\documentclass{article}
\usepackage{spconf,amsmath,graphicx,hyperref}

\usepackage{url}            
\usepackage{booktabs}       
\usepackage{amsfonts}       
\usepackage{nicefrac}       
\usepackage{microtype}      
\usepackage{acronym}
\usepackage{cite}
\usepackage{xcolor}         
\usepackage{algorithm}
\usepackage{algorithmic}
\usepackage{arydshln}
\usepackage{pifont}
\usepackage[normalem]{ulem}

\usepackage{amsmath,amsfonts,bm}

\def\1{\bm{1}}

\def\rvn{{\mathbf{n}}}

\def\rvp{{\mathbf{p}}}

\def\rvr{{\mathbf{r}}}
\def\rvs{{\mathbf{s}}}

\def\rvw{{\mathbf{w}}}
\def\rvx{{\mathbf{x}}}
\def\rvy{{\mathbf{y}}}
\def\rvz{{\mathbf{z}}}

\def\rmI{{\mathbf{I}}}

\DeclareMathAlphabet{\mathsfit}{\encodingdefault}{\sfdefault}{m}{sl}
\SetMathAlphabet{\mathsfit}{bold}{\encodingdefault}{\sfdefault}{bx}{n}

\def\gH{{\mathcal{H}}}

\def\gL{{\mathcal{L}}}

\def\gN{{\mathcal{N}}}

\def\sR{{\mathbb{R}}}

\newcommand{\xmark}{\ding{55}}%

\acrodef{DPS}{diffusion posterior sampling}
\acrodef{NMF}{non-negative matrix factorization}
\acrodef{EM}{expectation-maximisation}
\acrodef{SNR}{signal-to-noise ratio}
\acrodef{ODE}{ordinary differential equation}
\acrodef{STFT}{short-time Fourier Transform}
\acrodef{SNR}{signal-to-noise ratio}
\acrodef{SIR}{signal-to-interference ratio}
\acrodef{ESTOI}{extended short-term objective intelligibility}
\acrodef{PESQ}{perceptual evaluation of
speech quality}
\acrodef{SI-SDR}{scale-invariant signal-to-distortion rati}

\title{diffusion-based unsupervised audio-visual speech separation \\
in noisy environments with noise prior}
\name{Yochai Yemini\textsuperscript{1}, Rami Ben-Ari\textsuperscript{2}, Sharon Gannot\textsuperscript{1} and Ethan Fetaya\textsuperscript{1} 
}
\address{\textsuperscript{1}Faculty of Engineering, Bar-Ilan University, Ramat Gan, Israel \\
\textsuperscript{2}OriginAI   \\
\texttt{yochai.yemini@biu.ac.il}
}
\begin{document}
%
\maketitle
\begin{abstract}

In this paper, we address the problem of single-microphone speech separation in the presence of ambient noise. We propose a generative unsupervised technique that directly models both clean speech and structured noise components, training exclusively on these individual signals rather than noisy mixtures. Our approach leverages an audio-visual score model that incorporates visual cues to serve as a strong generative speech prior. By explicitly modelling the noise distribution alongside the speech distribution, we enable effective decomposition through the inverse problem paradigm. We perform speech separation by sampling from the posterior distributions via a reverse diffusion process, which directly estimates and removes the modelled noise component to recover clean constituent signals. Experimental results demonstrate promising performance, highlighting the effectiveness of our direct noise modelling approach in challenging acoustic environments.

\end{abstract}
\section{Introduction}
\label{sec:intro}
One of the fundamental challenges in speech processing is recovering clean speech from recordings conducted in a noisy environment. Over the last decade, this field has been dominated by neural network techniques. Most of them address the noise removal challenge in a supervised manner, i.e., a neural network is trained to predict clean speech from its noisy version. Although the supervised approach achieves excellent results for the task it was trained on, its main drawback is its inflexibility.
When new acoustic conditions are encountered, supervised methods have to retrain on the new scenario to avoid performance degradation.

In contrast, unsupervised speech processing takes a different approach. In particular, unsupervised generative paradigms train a prior speech model which learns the data distribution of \textit{clean} speech signals. The domain knowledge on clean speech is used by the prior model to estimate the target speech from its noisy observation. The main benefit of this method is its flexibility. As there is no need to train any dedicated  model for each degradation setup, any speech or noise model and any number of speakers can be used. This virtue enables handling various speech processing problems, e.g., speech enhancement and speaker separation, within a unified framework. 

We focus on unsupervised algorithms that use diffusion models \cite{ho2020denoising,song2021scorebased} as speech priors to estimate clean speech from corrupted recordings through an inverse problem framework. 
In the inverse problem formulation, the clean speech is sampled with the \textit{posterior} speech distribution, conditioned on the corrupted speech. This technique has been successfully applied to various speech processing tasks, such as \cite{buddy,moliner2023solving,moliner2024blind,arraydps}.

Nevertheless, the applicability of the inverse problem paradigm to ambient noise suppression has been underexplored. A recently proposed speech enhancement method \cite{nmf1,nmf2} models the noise covariance matrix using \ac{NMF}. It then performs \ac{EM} iterations that alternate between sampling the clean signal from the posterior distribution and estimating the \ac{NMF} matrices of the noise.

The spectral content of noise signals can vary greatly, presenting intricate distributions. We postulate that the expressiveness of the \ac{NMF} noise model may be limited, resulting in suboptimal performance. Consequently, to better capture the noise distribution, we propose to deploy a separate diffusion model as a noise prior. For the speech prior, we learn a powerful audio diffusion model enhanced by visual cues. 
In this study, we harness the \ac{DPS} framework \cite{dps} to demonstrate the effectiveness of our approach on the challenging task of single-microphone speech separation in the presence of ambient noise. Our \ac{DPS} formulation considers the noise signal as an additional source, and estimates it alongside the speech sources. 

We dub our method DAVSS-NM, which stands for Diffusion-Based Audio-Visual Speech Separation with Noise Modelling. The experimental results show that DAVSS-NM significantly diminishes the performance gap between inverse problem techniques and a state-of-the-art predictive baseline. Consequently, as an unsupervised technique, DAVSS-NM presents a good tradeoff between flexibility and performance. 


\section{Problem Formulation}
\label{sec:prob_form}
Let $\rvy$ be a mixture of $K$ clean speech signals $\{\rvx_i\}_{i=1}^K\in\sR^d$, and noise signals $\rvn,\rvz\in\sR^d$, recorded with a single microphone:
\begin{equation}
    \rvy = \sum_{i=1}^{K} \rvx_i + \rvn + \rvz.
\label{eq:meas}
\end{equation}
All the signals are represented in the time-domain and are statistically independent. While $\rvn$ is an ambient noise coming from a structured noise distribution, $\rvz \sim \gN\left(\mathbf{0}, \sigma_{\rvz}^2 \rmI\right)$ is an artificially added noise with an arbitrarily low variance. $\rvz$ is added to ensure non-zero probability across the entire space for mathematical tractability. We will use the concatenation of all speech sources denoted as $\rvx_{1:K}\in\sR^{Kd}$ instead of $\{\rvx_i\}_{i=1}^K$ throughout the paper.

\section{Background}
\label{sec:background}
\subsection{Score-Based Diffusion Models}
Diffusion generative models aim to sample from a data distribution $p_{\textrm{data}}(\rvx)$ by initially sampling from a known noise distribution, and then iteratively denoising the sample. At the end of the denoising process, a sample from $p_{data}(\rvx)$ is obtained. The denoising process can be formulated as an \ac{ODE} \cite{song2021scorebased}, which incorporates the score function of intermediate diffusion states. Karras et al. \cite{karras2022elucidating} suggested the following variance exploding \ac{ODE}:
\begin{equation}
    d\rvx^\tau = -\dot{\sigma}(\tau)\sigma(\tau)\nabla_{\rvx^\tau}\log p_\tau(\rvx^\tau)d\tau
\label{eq:ode}
\end{equation}
where $p_\tau(\rvx^\tau|\rvx^0) = \gN(\rvx^{0}, \sigma^2(\tau)\rmI)$, $\tau$ is the diffusion step, and the dot denotes a derivative with respect to $\tau$. The \ac{ODE} transports a noise sample from $\gN(\mathbf{0}, \sigma^2(T_{\max})\rmI)$ to a sample $\rvx^0$ from $p_{\textrm{data}}(\rvx)$. The values of $\sigma(\tau)$ are chosen such that $\sigma(\tau)$ is monotonically increasing. In \cite{karras2022elucidating}, $\sigma(\tau)=\tau$ is chosen, and we follow this choice in our derivations. To ensure numerical stability, the smallest diffusion step is $\tau=T_{\min}$ and not $\tau=0$, for some arbitrarily small $T_{\min}$.

The score function $\nabla_{\rvx^\tau}\log p(\rvx^\tau)$ is intractable at inference time, because $\rvx^0$ is unknown. However, it can be approximated with a diffusion denoiser $D_{\theta}(\rvx^\tau, \tau)$ which is trained to output $\hat{\rvx}^0 \approx \rvx^0$ using the following $L_2$ loss:
\begin{equation}
    \mathbb{E}_{\tau, \rvx\sim p_{\textrm{data}}, \boldsymbol{\epsilon}\sim\gN(\mathbf{0},\rmI)}\left\|D_{\theta}(\rvx+\sigma(\tau)\boldsymbol{\epsilon},\tau) - \rvx\right\|^2.
\end{equation}
Finally, using Tweedie's Formula \cite{tweedies}, the score function can be estimated as:
\begin{equation}
    \nabla_{\rvx^\tau}\log p(\rvx^\tau) \approx \rvs_{\theta}(\rvx^\tau, \tau) = \frac{D_{\theta}(\rvx^\tau, \tau) - \rvx^\tau}{\sigma^2(\tau)}.
\end{equation}

\subsection{Diffusion Posterior Sampling}
The sampling procedure elaborated above describes an unconditional diffusion-based sampling procedure. This pretrained diffusion model can be utilised to estimate $\rvx$ when a degraded observation $\rvr=\gH(\rvx)+\rvw$ is available, where $\gH(\cdot)$ is a corruption operator and $\rvw \sim \gN(\mathbf{0}, \sigma_\rvw^2\rmI)$ is a noise term. The inverse problem framework suggests to estimate $\rvx$ by sampling from $p(\rvx|\rvr)$. This is implemented by replacing the prior $p_\tau(\rvx^\tau)$ in the reverse \ac{ODE} (\ref{eq:ode}) with $p_\tau(\rvx^\tau|\rvr)$. According to the Bayes' Theorem, $p_\tau(\rvx^\tau|\rvr) \propto p_\tau(\rvr|\rvx^\tau) p_\tau(\rvx^\tau)$, so the posterior sampling \ac{ODE} is:
\begin{equation}
    d\rvx^\tau = -[\nabla_{\rvx^\tau}\log p_\tau(\rvr|\rvx^\tau) + \nabla_{\rvx^\tau}\log p_\tau(\rvx^\tau)]\tau d\tau.
\label{eq:pos_ode1}
\end{equation}
The prior term can be readily estimated using $\rvs_\theta(\rvx^\tau, \tau)$. However, the likelihood score $p_\tau(\rvr|\rvx^\tau)$ is intractable. The \ac{DPS} algorithm \cite{dps} proposed to approximate the likelihood term by replacing the condition $\rvx^\tau$ with $\hat{\rvx}^0 = D_{\theta}(\rvx^\tau, \tau)$, i.e. a single-step estimation of the diffusion terminal state $\rvx^0$. Finally, since $\rvw$ is a Gaussian noise, \eqref{eq:pos_ode1} can be written as:
\begin{equation}
    d\rvx^\tau = -\left[\nabla_{\rvx^\tau}\log p_\tau(\rvx^\tau) - \zeta\nabla_{\rvx^\tau}||\rvr-\gH(\hat{\rvx}^0)||_2^2\right]\tau d\tau
\label{eq:pos_ode2}
\end{equation}
where $\zeta$ is a hyperparamter used to improve sampling quality, as proposed in \cite{dps}.

\section{Method}
\label{sec:method}
Given the mixture signal $\rvy$, we propose to recover the mixture components by sampling from the following posterior distribution:
\begin{equation}
    p(\rvx_{1:K}, \rvn | \rvy) \propto p(\rvy|\rvx_{1:K}, \rvn)p(\rvn)\prod_{i=1}^K p(\rvx_i)
\label{eq:posterior}
\end{equation}
where we used the statistical independence of the sources. 

We now wish to harness the inverse problem framework by utilising the diffusion reverse process to sample $\rvx_{1:K}$ and $\rvn$ from the posterior. To this end, due to the independence of the sources it stems that the following $K+1$ \acp{ODE} need to be solved for $i=1,\ldots,K$:
\begin{subequations}
\begin{align}
    &d\rvx_i^\tau = -[\nabla_{\rvx_i^\tau} \log p(\rvy|\rvx_{1:K}^\tau, \rvn^\tau) + \nabla_{\rvx_i^\tau} \log p_\tau(\rvx_i^\tau)]\cdot \tau d\tau 
    \label{eq:ode_speech}
    \\
    &d\rvn^\tau = -[\nabla_{\rvn^\tau} \log p(\rvy|\rvx_{1:K}^\tau, \rvn^\tau) + \nabla_{\rvn^\tau} \log p_\tau(\rvn^\tau)]\cdot \tau d\tau
\label{eq:ode_noise}
\end{align}
\end{subequations}

For computing the prior terms in \eqref{eq:ode_speech} and \eqref{eq:ode_noise}, we use two separate diffusion models defined on the clean speech and noise priors, respectively. The noise prior is represented by the score function $\rvs_{\phi}(\rvn^\tau, \tau)$ and the diffusion denoiser $G_{\phi}^N(\rvn^\tau,\tau)$. For the speech prior, we assume the availability of visual features $V_i$, extracted from a lip video corresponding to each speaker. The visual modality can be beneficial to improve the accuracy of the score estimation. Consequently, an audio-visual score function $\rvs_{\theta}(\rvx_i^\tau, \tau, V_i)$ and the denoiser $F_\theta^S(\rvx_i^\tau,\tau,V_i)$ coupled with classifier-free guidance \cite{ho2021classifierfree} are used. Note that all the speech signals share the same speech prior. 

Since the likelihood terms in (\ref{eq:ode_speech}) and (\ref{eq:ode_noise}) are conditioned on $\rvx_{1:K}^\tau$ and $\rvn^\tau$, they are intractable.
We thus rely on the \ac{DPS} algorithm \cite{dps}, and extend its original form from single to multiple signals and two different prior distributions, i.e. speech and ambient noise. 
Similarly to \eqref{eq:pos_ode2}, it stems from the Gaussianity of $\rvz$ that the likelihood score terms become:
\begin{subequations}
\begin{align}
    &\nabla_{\rvx_i^\tau} \log p(\rvy|\hat{\rvx}_{1:K}^0, \hat{\rvn}^0) = 
    -\zeta_{\rvx}(\tau)\nabla{\rvx_i^\tau}\gL_{\text{rec}}^{t} 
    \\
    &\nabla_{\rvn^\tau} \log p(\rvy|\hat{\rvx}_{1:K}^0,\hat{\rvn}^0) = 
    -\zeta_{\rvn}(\tau)\nabla{\rvn^\tau}\gL_{\text{rec}}^{t}.
\end{align}
\end{subequations}
where $\gL_{\text{rec}}^\tau = ||\rvy - \sum_i \hat{\rvx}_i^0 - \hat{\rvn}^0||_2^2$ is the mixture reconstruction loss in the time domain, and $\zeta_{\rvx}(\tau), \zeta_{\rvn}(\tau)$ are normalisation scalars. 
To achieve more accurate gradients, we follow \cite{buddy} by replacing the time-domain mixture reconstruction error with its frequency-domain counterpart:
\begin{equation}
\gL_{\text{rec}}^{f} = 
\left\|S(\rvy) - S\left(\sum_{i=1}^K \hat{\rvx}_i^0 + \hat{\rvn}^0)\right)\right\|_2^2
\end{equation}
where $S(\rvy)=|\textrm{STFT}(\rvy)|^{\frac{2}{3}}\exp{j\angle \textrm{STFT}(\rvy)}$, and $\textrm{STFT}(\cdot)$ is the \ac{STFT}.

We follow \cite{moliner2023solving,moliner2024blind} by using $\zeta_{\rvx}(\tau), \zeta_{\rvn}(\tau)$ which contribute to the sampling stability:
\begin{subequations}
\begin{align}
    &\zeta_{\rvx}(\tau) = \frac{\zeta \sqrt{d}}{\tau\left\|\nabla_{\rvx_{1:K}^\tau}\gL_{\text{rec}}^{f}\right\|_2}
    \\
    &\zeta_{\rvn}(\tau) = \frac{\zeta \sqrt{d}}{\tau\left\|\nabla_{\rvn^\tau}\gL_{\text{rec}}^{f}\right\|_2}.
\end{align}
\end{subequations}
We summarise in Algorithm \ref{alg:post_score} the steps for obtaining the posterior score estimations $\{\rvp_{\rvx_i^\tau}\}_{i=1}^K, \rvp_{\rvn^\tau}$ of the speech and noise components.
Finally, the \acp{ODE} in (\ref{eq:ode_speech}), (\ref{eq:ode_noise}) are solved with the stochastic second order sampler proposed in \cite{karras2022elucidating}.

\begin{algorithm}[t]
\caption{Posterior Score Estimation}
\begin{algorithmic}[1]
\REQUIRE $\rvy, \rvx_{1:K}^\tau, \rvn^\tau, \tau, \zeta, d, F_{\theta}^S, G_{\phi}^N$
\STATE $\hat{\rvn}^0 \leftarrow G_{\phi}^N(\rvn^\tau, \tau)$
\STATE $\rvs_{\phi}(\rvn^\tau, \tau) \leftarrow \frac{\hat{\rvn}^0-\rvn^\tau}{\tau^2}$  \hfill $\triangleright$ Noise prior score
\FOR{$i = 1, \ldots, K$}
    \STATE $\hat{\rvx}_i^0 \leftarrow F_{\theta}^S(\rvx_i,\tau,V_i)$
    \STATE $\rvs_\theta(\rvx_i^\tau, \tau) \leftarrow \frac{\hat{\rvx}_i^0-\rvx_i^\tau}{\tau^2}$ \hfill $\triangleright$ Speech prior score
\ENDFOR
\STATE $\gL_{\text{rec}}^f \leftarrow \left\|S(\rvy) - S(\sum_{i=1}^K\hat{\rvx}_i^0 + \hat{\rvn}^0)\right\|_2^2$
\STATE $\zeta_{\rvn}(\tau) = \frac{\zeta \sqrt{d}}{\tau\left\|\nabla_{\rvn^\tau}\gL_{\text{rec}}^f\right\|_2}$
\STATE $\rvp_{\rvn^\tau} \leftarrow \rvs_{\phi}(\rvn^\tau, \tau)-\zeta_{\rvn}(\tau)\nabla_{\rvn^\tau}\gL_{\text{rec}}^f$  \hfill $\triangleright$ Noise posterior score
\STATE $\zeta_{\rvx}(\tau) = \frac{\zeta \sqrt{d}}{t\left\|\nabla_{\rvx_{1:K}^\tau}\gL_{\text{rec}}^f\right\|_2}$
\FOR{$i = 1, \ldots, K$}
    \STATE $\rvp_{\rvx_i^\tau} \leftarrow \rvs_\theta(\rvx_i^\tau, \tau)-\zeta_{\rvx}(\tau)\nabla_{\rvx_i^\tau}\gL_{\text{rec}}^f$  \hfill $\triangleright$ Speech posterior score
\ENDFOR
\STATE \textbf{return} $\{\rvp_{\rvx_i^\tau}\}_{i=1}^K, \rvp_{\rvn^\tau}$
\end{algorithmic}
\label{alg:post_score}
\end{algorithm}

\begin{table*}[t]
\centering
\caption{Speech separation results of DAVSS-NM and the baselines. Best results are in bold, second best are underlined. 
}
\label{tab:results}
\begin{tabular}{l c c c c c c c c}
\toprule
Method & Trained noise & Unupervised &
\multicolumn{3}{c}{VoxCeleb2 + DNS} &
\multicolumn{3}{c}{VoxCeleb2 + WHAM!} \\
\cmidrule(lr){4-6} \cmidrule(lr){7-9}
 & & & SI-SDR $\uparrow$ & PESQ $\uparrow$ & ESTOI $\uparrow$ & SI-SDR $\uparrow$ & PESQ $\uparrow$ & ESTOI $\uparrow$ \\
\midrule
Input & - & - & -2.66 & 1.46 & 0.38 & -2.62 & 1.52 & 0.36 \\
\midrule
FlowAVSE\cite{flowavse} & DNS & \xmark & \textbf{7.82} & \textbf{2.15} & \textbf{0.65} & \textbf{7.13} & \textbf{2.18} & \textbf{0.62} \\
\hdashline
VisualVoice\cite{gao2021VisualVoice} & DNS & \xmark & 1.89 & 1.85 & 0.52 & 1.45 & 1.86 & 0.48 \\
AV-UDiffSE\cite{nmf2} & DNS & \checkmark & -2.33 & 1.76 & 0.44 & -2.27 & 1.81 & 0.44 \\
DAVSS-NM (Ours) & DNS & \checkmark & \underline{5.06} & \underline{2.06} & \underline{0.61} & \underline{4.54} & \underline{2.04} & \underline{0.58} \\
\midrule
FlowAVSE\cite{flowavse} & WHAM! & \xmark & \textbf{7.04} & \textbf{2.19} & \textbf{0.62} & \textbf{6.41} & \textbf{2.07} & \textbf{0.61} \\
\hdashline
VisualVoice\cite{gao2021VisualVoice} & WHAM! & \xmark & 0.03 & 1.67 & 0.46 & 0.35 & 1.78 & 0.46 \\
AV-UDiffSE\cite{nmf2} & WHAM! & \checkmark & -2.27 & 1.81 & 0.44 & -2.33 & 1.76 & 0.44 \\
DAVSS-NM (Ours) & WHAM! & \checkmark & \underline{4.58} & \underline{2.07} & \underline{0.58} & \underline{3.93} & \underline{1.95} & \underline{0.58} \\
\bottomrule
\end{tabular}
\end{table*}

\section{Experimental Study}
\label{sec:results}
In this section, we present the experimental study used to evaluate the proposed method in comparison with baseline approaches. We also describe the datasets and provide key implementation details.

\noindent \textbf{Data. } To align with the dataset our visual encoder was trained on, the English-only version of the audio-visual dataset VoxCeleb2 \cite{vox2} was used as the source of clean speech utterances. The English-only videos were curated by \cite{avhubert}, amounting to a total length of 1,759 hours. The full-face videos are converted to mouth-region videos by following the prescription in \cite{braven}. The sample rates of the audio and video are 16 KHz and 25 Hz, respectively.

For noise recordings, two datasets are used. The first is WHAM! \cite{wham}, comprising roughly 30h/10h/5h of background noise for training/validation/test. All noise signals were recorded in urban environments such as parks, restaurants, and office buildings. The second dataset is DNS \cite{dns}. It consists of about 65,000 clips belonging to 150 noise classes collected from AudioSet \cite{audioset} and Freesound\footnote{\url{https://freesound.org/}}. The total duration of the recordings is 181 hours. We randomly choose 64,000 noise signals for training, 100 for validation and 900 for evaluation.

\noindent \textbf{Networks architecture. }$F_\theta^S$ and $G_\phi^N$ are based on NCSN++M \cite{storm}, a light U-Net architecture. Similarly to \cite{buddy}, the NCSN++M backbone is wrapped with STFT and inverse STFT. 
All \ac{STFT} operations use Hann window of 510 samples and hop length of 160. This results in 256 frequency bins.

For $G_\phi^N$, we increase the number of residual layers to 2. For $F_\theta^S$, the NCSN++M backbone is augmented with visual features. The lip video frames are encoded using BRAVEn \cite{braven} finetuned on visual speech recognition. Each frame is encoded to a feature vector in $\sR^{1024}$. The visual modality is integrated with the NCSN++M at resolutions 64, 32 and the bottleneck, by first matching the visual resolution to the audio resolution and then modulating the audio features via FiLM layers \cite{film}. Altogether, our $F_\theta^S$ and $G_\phi^N$ have 129.5M and 39.7M trainable parameters, respectively.

\noindent \textbf{Baselines. }
Three baselines are used, which we train from scratch. The first is VisualVoice \cite{gao2021VisualVoice}, a supervised audio-visual discriminative model which encourages speaker identity consistency between the separated speech and its corresponding video. Another baseline is FlowAVSE \cite{flowavse}, a {\it supervised} audio-visual generative model based on flow matching. These supervised algorithms are trained with the speech-noise combinations VoxCeleb2+WHAM! and VoxCeleb2+DNS. We demonstarte our method on mixtures of two speakers. The \ac{SIR} and \ac{SNR} during training matched the evaluation protocol described in the sequel.

The last baseline AV-UDiffSE \cite{nmf2} is an {\it unsupervised} generative method based on the inverse problems methodology. As it was originally designed for speech enhancement with a single speaker, we modified it to conform to our scenario by  adapting the derivations in \cite{nmf2} to a multiple-speaker case.
Additionally, we replace the audio-based backbone with our audio-visual architecture.

\noindent \textbf{Training setup. }
To train $F_\theta^S, G_\phi^N$, 4s audio segments were sampled, resulting in $d=64,000$ and 400 \ac{STFT} frames. For $F_\theta^S$, it translates to 100 video frames. We use the discretised diffusion noise shedule $\sigma(\tau)$ from \cite{karras2022elucidating}
with $\rho=10$. During training, $T_{\max}=10, T_{\min}=1e-5$ are used.
The batch size for training $F_\theta^S, G_\phi^N$ is 16 for all experiments, and we use the Adam optimiser with learning rate 1e-4.

\noindent \textbf{Posterior sampling. }
At inference time, the second-order sampler from \cite{karras2022elucidating} is used. We choose $T_{\max}=4,T_{\min}=1e-5,S_{\textrm{churn}}=30,\zeta=0.5$, and discretise the \ac{ODE} (\ref{eq:ode_speech}), (\ref{eq:ode_noise}) with 400 diffusion timesteps. A classifier-free guidance weight of 0.5 is used for WHAM! noise, and 0.8 for DNS. Unlike previous methods \cite{moliner2024blind,buddy,nmf1} which use warm initialisation, our initial diffusion states $\rvx_i^{T_{\max}}, \rvn^{T_{\max}}$ are set to have zero mean. A zero-mean initial state is also used for our implementation of AV-UDiffSE, and the number of \ac{EM} iterations is set to 9.

\noindent \textbf{Evalutation protocol. }
The evaluation set comprised mixtures of two speakers and a background noise signal. The \ac{SIR} values were -5/0/5 dB, and the \ac{SNR} was randomly sampled from the range [-3,3], defined with respect to the low-energy speaker. For each \ac{SIR} value, 500 mixtures were generated. We examined two scenarios, matched and unmatched. In the matched setup, the methods were tested on the speech-noise datasets on which they were trained. In the unmatched regime, a model trained on WHAM! noise was tested on DNS noise and vice versa. This cross-domain test allows us to evaluate how well the models generalise. We used the \ac{PESQ} score, \ac{ESTOI}, and \ac{SI-SDR} measured in dB to assess performance.

\noindent \textbf{Discussion.}
Table \ref{tab:results} presents the performance of DAVSS-NM and the baselines. It is evident that the supervised FlowAVSE is top performing. For all speech-noise combinations, DAVSS-NM is the runner-up and best  unsupervised method. It improved the metrics over AV-DiffSE, thus substantially narrowing the gap between unsupervised and supervised paradigms. DAVSS-NM even outperforms VisualVoice, which is a supervised approach. We suppose that VisualVoice did not obtain better results because it is based on \ac{STFT} masking, which is adversely affected by our challenging scenario.

In the unmatched regime, when trained on WHAM! noise, all methods except for VisualVoice relatively maintained performance. Note that since AV-UDiffSE formulation does not require training on the noise data, its results can be copied from the relevant entries. Surprisingly, when trained on DNS noise, all methods achieved performance superior to their counterparts trained on WHAM!. We postulate that this phenomenon occurred because DNS is a highly diverse noise dataset, whose distribution likely covers that of WHAM!. 

\vspace{-7pt}
\section{Conclusion}
\label{sec:conclusion}
\vspace{-7pt}
We presented a joint unsupervised diffusion-based, speech-separation and ambient noise removal approach, leveraging a noise distribution modelling. It was developed under the inverse problem framework by coupling the noise prior with an audio-visual speech prior.  Our method, DAVSS-NM, has two main virtues. Firstly, unlike supervised methods which require retraining when the degradation model changes, DAVSS-NM does not rely on paired training data, making it highly adaptable. Secondly, our evaluations show that our approach considerably reduces the performance gap between unsupervised techniques and a state-of-the-art supervised algorithm. 




\bibliographystyle{IEEEbib}
\bibliography{strings,refs}

\end{document}